\documentclass[useAMS,usenatbib]{mn2e}

\usepackage{graphicx}

\title[Tidal disruption of NEAs - a case of P\v{r}\'{i}bram]
  {Tidal disruption of NEAs
   - a case of P\v{r}\'{i}bram meteorite}
\author[J. T\'{o}th et al.]
  {J.~T\'{o}th,$^1$
  P.~Vere\v{s},$^1$ and L.~Korno\v{s}$^1$ \\
  $^1$Faculty of Mathematics, Physics, and Informatics, Comenius University, 842 48 Bratislava, Slovakia
      Mlynsk\'{a} dolina\\
 }
\date{Released 2002 Xxxxx XX}

\pagerange{\pageref{firstpage}--\pageref{lastpage}} \pubyear{2002}

\def\LaTeX{L\kern-.36em\raise.3ex\hbox{a}\kern-.15em
    T\kern-.1667em\lower.7ex\hbox{E}\kern-.125emX}

\begin{document}

\label{firstpage}

 \maketitle

\begin{abstract}

This work studies the dynamical evolution of a possible meteor
stream along the orbit of the P\v{r}\'{i}bram meteorite, which
originated in the tidal disruption of the putative
rubble-pile-like parent body during a close approach to the Earth.
We assumed the disruption at the time when the ascending or
descending node of the parent orbit was close to the Earth's
orbit. In the last 5000 years, the P\v{r}\'{i}bram orbit has
crossed the Earth orbit twice. It happened about 4200 years and
3300 years ago. In both cases, we modeled the release of particles
from the simplified model of rotating asteroid, and traced their
individual orbital evolution to the current date. It takes several
hundred years to spread released meteoroids along the entire orbit
of the parent body. Even today, the stream would be relatively
narrow.

Considering a model parent body with physical parameters of the
asteroid Itokawa, the complete disintegration of the object
produced 3.8$\times10^{11}$ meteoroid particles with diameter
$\geq$ 1\,cm. The meteor activity observed from the Earth is
revealed and justification of follow-up observation during
suggested activity of the shower in the first two weeks of April
is discussed.

The Earth's tidal forces would disintegrate a fraction of NEA
population into smaller objects. We evaluate the upper limit of
mass of disintegrated asteroids within the mean NEA lifetime and
the contribution of disrupted matter to the size distribution of
the NEA.

\end{abstract}


\section{Introduction}

The idea of meteoroid streams coming from asteroids was presented
in the past (\citealt{oli}, \citealt{hof}, \citealt{hal},
\citealt{por}, \citealt{trigo}, etc.). For instance \citet{hal}
analyzed orbits of 89 bolides from the Canadian and American
bolide network (MORP and Prairie Network) that might survive the
atmospheric flight with a non-zero remaining mass and suggested
the existence of streams producing meteorites. The mentioned
authors concluded that these streams originated from asteroids.
\citet{por} searched for genetic relations between asteroids and
bolide meteoroid streams. The authors investigated the evolution
of the orbits and only include as real asteroid-stream pairs those
where the evolution was also similar over 5000 years.

The question is what could cause the escape of the matter from the
parent asteroids if not collision with another cosmic body? There
are several other mechanisms that may cause the disintegration of
the asteroidal body: YORP effect spin-up, thermal stress break-up
and tidal disruption of the asteroid during its close fly-by
around the planet. Among the terrestrial planets the Earth is the
dominant planet that is able to disrupt and distort approaching
bodies. Unlike in the case of comets, when the stream of
meteoroids is regularly replenished, asteroids can undergo the
mentioned events sparsely and streams of meteoroids are rather
created by a single event. Therefore, the spatial density of such
a stream should be lower and the expected activity hardly
distinguished from the sporadic background.

The major motivation for this work was the fall of the
Neuschwanstein meteorite on April 6, 2002. The analysis of its
heliocentric orbit revealed that the orbit was almost identical to
the orbit of P\v{r}\'{i}bram meteorite observed on April 7, 1959
(\citealt{spur}). Although the two meteorites are of different
types (P\v{r}\'{i}bram an ordinary H5 chondrite, Neuschwanstein an
enstatite EL6 chondrite) and their cosmic ray exposure times
differ (12 and 48 million years, respectively), \citet{spur}
proposed an existence of meteoroid stream that might originated in
the tidal break-up of an heterogeneous rubble-pile-like asteroid.
The evidence of high internal porosity of asteroids, e.g. (253)
Mathilde, irregular shapes, e.g. (1620) Geographos, (216)
Kleopatra, (66391) 1999\,KW$_{4}$, detailed surface images of
(25134) Itokawa and spin barrier of asteroids with diameters
exceeding approx. 200\,m imply that a significant fraction of
asteroids could have heavily cracked interiors or rubble-pile-like
structures. Such conglomerates hold together only by a relatively
weak gravity. Surface images of Itokawa also suggested a movement
of rubble and dust on the surface in the past (\citealt{miy a},
\citealt{miy b}) and further theoretical works (\citealt{rich},
\citealt{sch}, \citealt{ross}) present that asteroids may change
their shapes or lose mass by the YORP spin-up or tidal break-up.
If only surface is resurfaced, some boulders may hide and some
emerge, which could explain different cosmic ray exposure times
for meteoroids coming from the same parent body as seen in the
case of P\v{r}\'{i}bram and Neuschwanstein.

Our previous paper (\citealt{kor1}) showed that the orbital
evolution of P\v{r}\'{i}bram and Neuschwanstein exhibits similar
behavior at least during the last 5000 years. The
Southworth-Hawkins D-criterion for both orbits remains lower than
$D_{SH}<0.07$ and the difference between longitudes of perihelion
less than $3\degr$. Moreover, cloned orbits derived within the
orbit uncertainties of both meteorites are stable as well.
Therefore, the putative stream of meteoroids along the orbit of
P\v{r}\'{i}bram could be stable at least for thousands of years,
which is consistent with \citet{pg} who derived its decoherence
time of about 50\,000 years. If there is such a stream, it must
have originated relatively recently.

The search for possible members of the P\v{r}\'{i}bram meteor
stream brought several suspicious meteors from the IAU meteor
orbit database and asteroids from the Minor Planet Center database
(\citealt{spur}, \citealt{kor1}). For instance the asteroid 2002
QG$_{46}$ and meteor 161E1 are relatively near to P\v{r}\'{i}bram
orbit, but they do not exhibit orbit similarity as P\v{r}\'{i}bram
and Neuschwanstein. Even though \citet{pg} showed that
statistically such a close pair could exist as a coincidence. Such
conclusion shall not be drawn (\citealt{kor2}) while the count
estimates of one meter size NEAs differ more than two order of
magnitude.

In this paper, we study the orbital evolution of potential
meteoroid stream along the orbit of P\v{r}\'{i}bram. The stream
originated in the tidal disruption of the parent body in the past,
several thousand years ago. The second and third chapters deal
with the origin and particle motion of the meteoroid stream, its
orbital evolution and the activity of such a stream today. The
fourth chapter deals with the fly-by frequency of the NEAs within
the Earth Roche limit. Available Earth impact frequency is
adapted. Moreover, it is assumed that the limit rises as a
function of the spin rate of the asteroid and the cross-section
target plane of the asteroid disruption grows. The fifth chapter
evaluates the fraction of NEAs that could be tidally disrupted by
the Earth during the median population lifetime in the NEO space,
and the amount of disintegrated matter and size-frequency of the
altered population is discussed.

\section{Meteoroid stream as a result of asteroid tidal break-up}

The complex problem of the asteroid break-up due to the tidal
forces has been presented in several papers (\citealt{bott1997},
\citealt{bott1998}, \citealt{rich}, \citealt{shar},
\citealt{hols}). According to the recent results a weak
rubble-pile asteroid is deformed during a close fly-by around a
planet (Earth) to the shape of the rotational ellipsoid and the
matter starts to leave the surface from the ends of the longest
axis. Even if the asteroid survives the close encounter, a
significant amount of matter leaves the surface in form of
regolith, pebbles and boulders.

In the work of \citet{kor2}, we modeled the release of meteoroid
particles from the parent asteroid during its fly-by around the
Earth within the Roche limit. The parent body was placed on the
orbit of P\v{r}\'{i}bram meteorite. Particles loosed from the
surface reached escape velocities around 10 $cm\,s^{-1}$. In the
model of hyperbolic motion of the progenitor asteroid around the
Earth, modeled particles are only several hundred meters away from
the asteroid when the geocentric distance reaches 100\,000 km
after the perigee passage. The orbital evolution of loose
particles only weekly depends on the progenitor's orbit. After
hundreds of years particles are distributed along the entire orbit
of the parent and even after 1000 years ($\sim$\,250 revolutions
around the Sun) orbital elements of individual particles are still
not much dispersed.

\begin {figure}
 \includegraphics[width=8.2cm,angle=0]{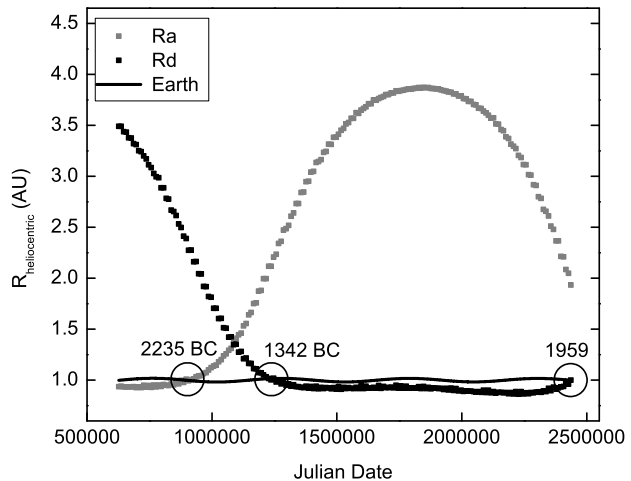}
 \caption{Evolution of heliocentric distance of ascending and descending nodes of
  P\v{r}\'{i}bram and Neuschwanstein during the last 5000 years.}
  \label{F1}
\end{figure}

According to the orbital evolution of P\v{r}\'{i}bram and
Neuschwanstein, we investigated the heliocentric distance of the
ascending and descending nodes (Figure 1). During the last 5000
years, the node has gotten close to the Earth's orbit twice. We
assigned the tidal disruption event to these dates.  The period
when the node got close to the Earth orbit takes tens of years. We
selected the middle of the time interval as the exact date of the
event. For both dates, we performed simplified tidal disruption of
the modeled Itokawa like asteroid as in the previous work
(\citealt{kor2}). During each event, 3100 particles were
numerically integrated and the orbital evolution was traced.
Position and velocity vectors of released particles with respect
to the parent asteroid are displayed in Figure 2.

\begin{figure}
 \includegraphics[width=6.2cm,angle=-90]{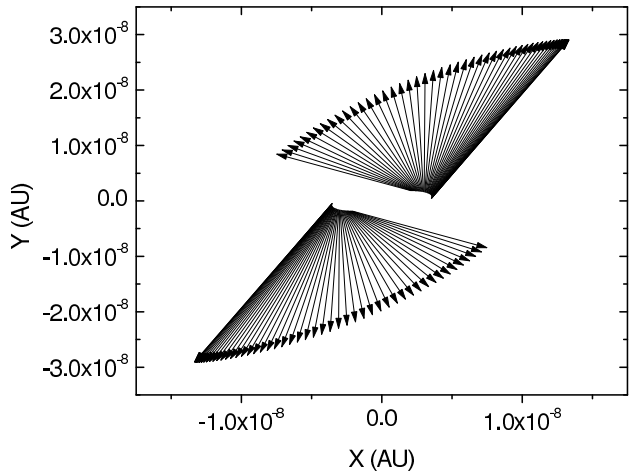}
 \caption{Distribution of position and velocity
  vectors of particles leaving the surface of the asteroid after its
  tidal disruption with respect to the center of the mass of the
  asteroid. The figure depicts the situation in the distance of
  100\,000 km from the Earth after the perigee and Roche limit
  passage.}
  \label{F2}
\end{figure}

The older event occurred $\sim$\,4200 years ago (JD\,=\,905\,120),
the more recent event 3300 years ago (JD\,=\,1\,231\,420). In the
study, the multistep Adams-Bashforth-Moulton type up to the 12th
order numerical integrator, with a variable step-width, was used.
All planets were considered as perturbing bodies and Earth and
Moon were treated separately.

\section{Meteoroid stream characteristics}

It takes several centuries to redistribute all released particles
along the entire orbit. Even today, 3300 years (4200 years,
respectively) after the event, particles are orbiting the Sun in a
narrow stream along the orbit of the progenitor (Figure 3). The
older stream is spread more widely because the node of the orbit
crosses the Earth's orbit about 1000 years later after the
release. That causes additional perturbation of meteoroid orbits.

\begin{figure}
\includegraphics[width=5.8cm,angle=-90]{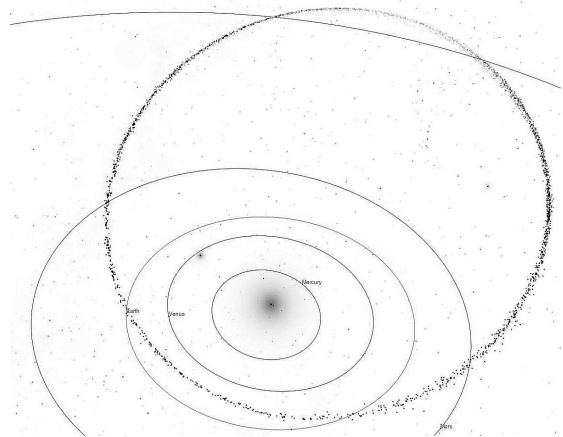}
 \caption{The distribution of particles created by
  the tidal break-up of the parent asteroid. Modeled particles are
  spread along the progenitor's orbit within several hundred years.}
  \label{F3}
\end{figure}

The distribution of orbital elements to the current date implies
that particles created in both events are on stable orbits (Figure
4). The dispersion in the semimajor axis does not exceed 0.05\,AU,
0.04 in the eccentricity, $5\degr$ in the inclination for the
older event and  $2\degr$ for the younger event. Argument of
perihelion and longitude of ascending node show higher dispersion;
however, their sum, the longitude of the perihelion, is within
several degrees. The exact elements of P\v{r}\'{i}bram and
Neuschwanstein depicted in the Figure 4 are within the intervals
of stream orbital elements.

\begin{figure*}
\centerline{\includegraphics[width=4.5cm,angle=-90]{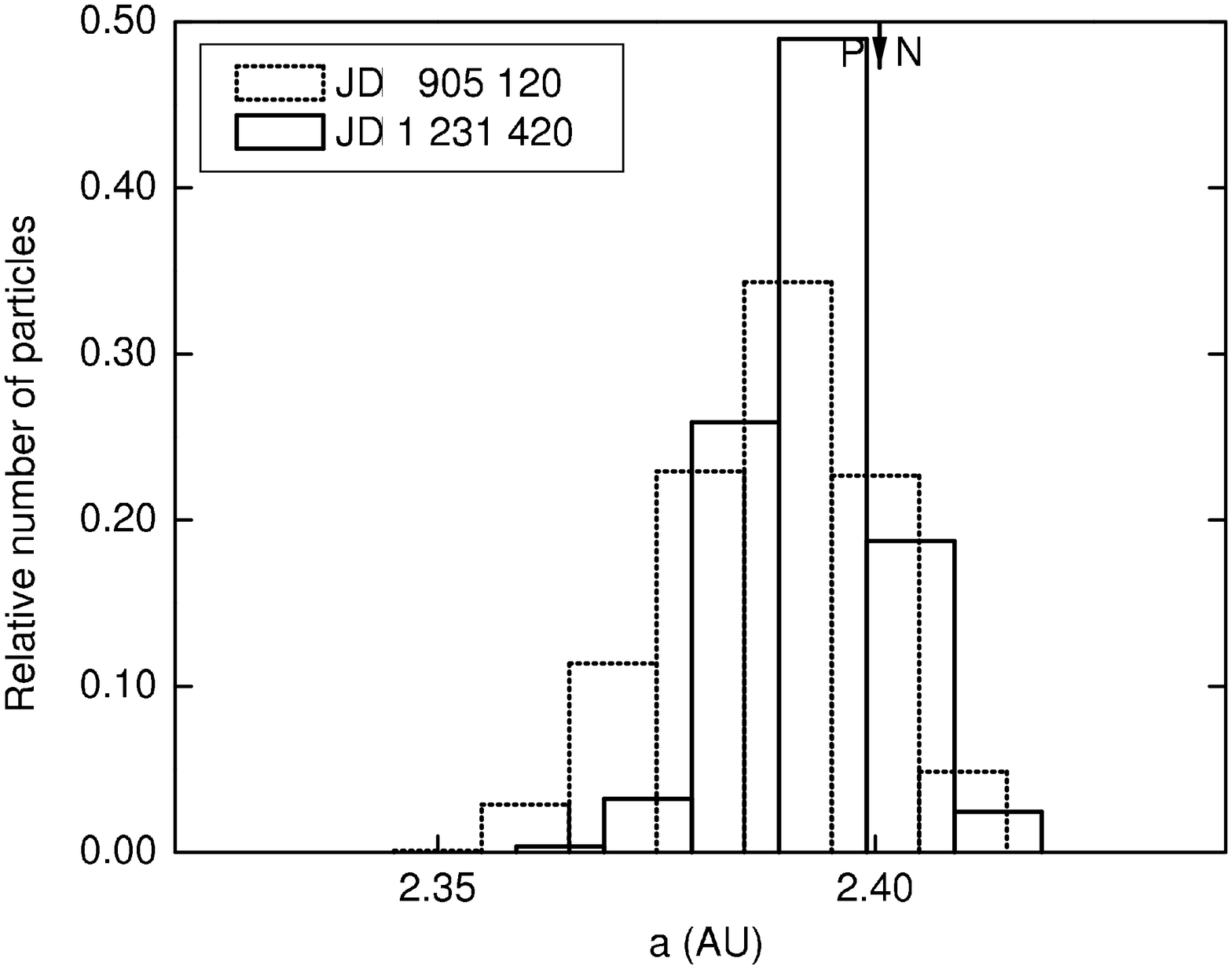}
    \hspace{0.3cm}
            \includegraphics[width=4.4cm,angle=-90]{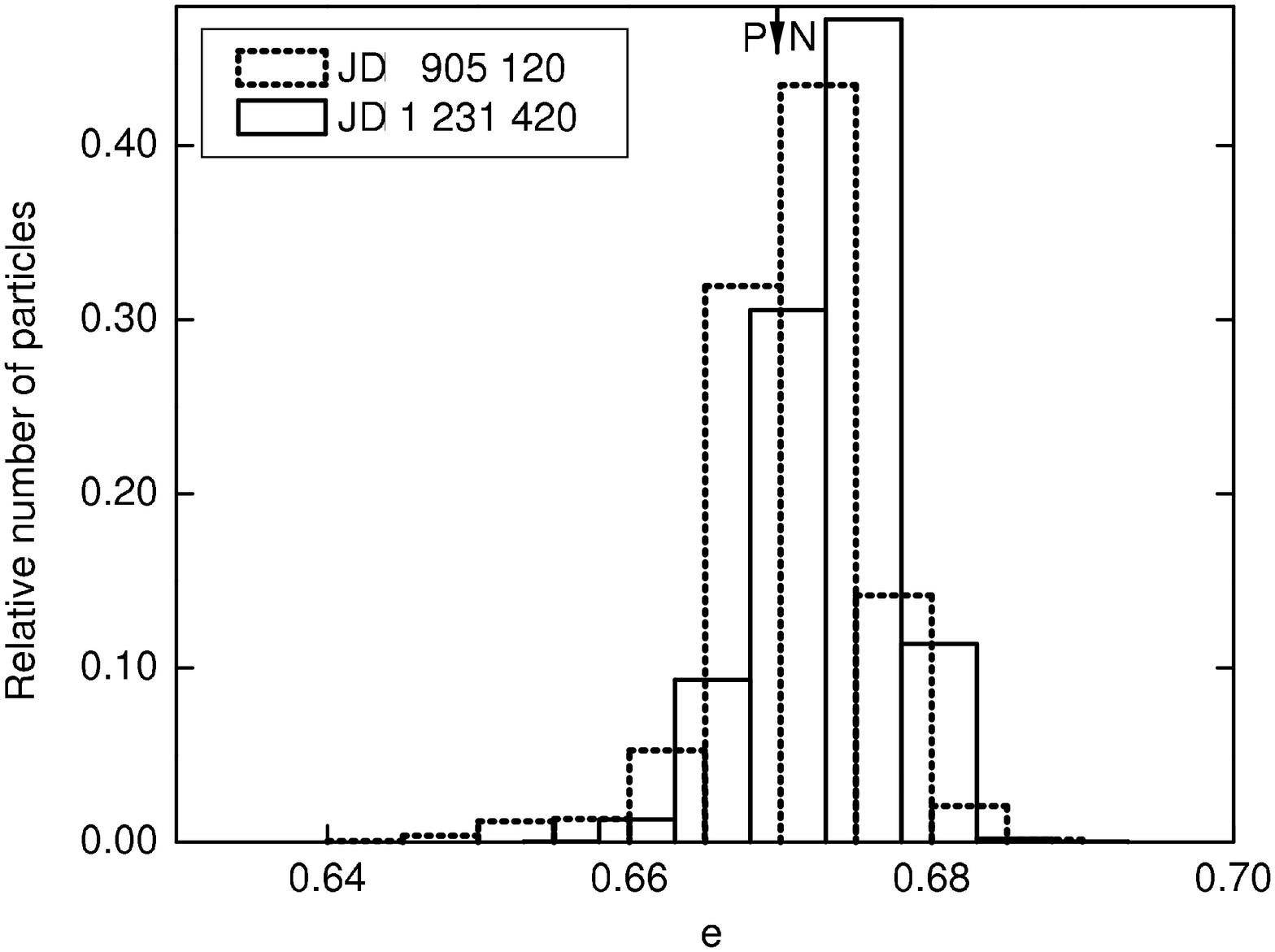}}
 \vspace{0.7cm}
\centerline{\includegraphics[width=5.6cm,angle=0]{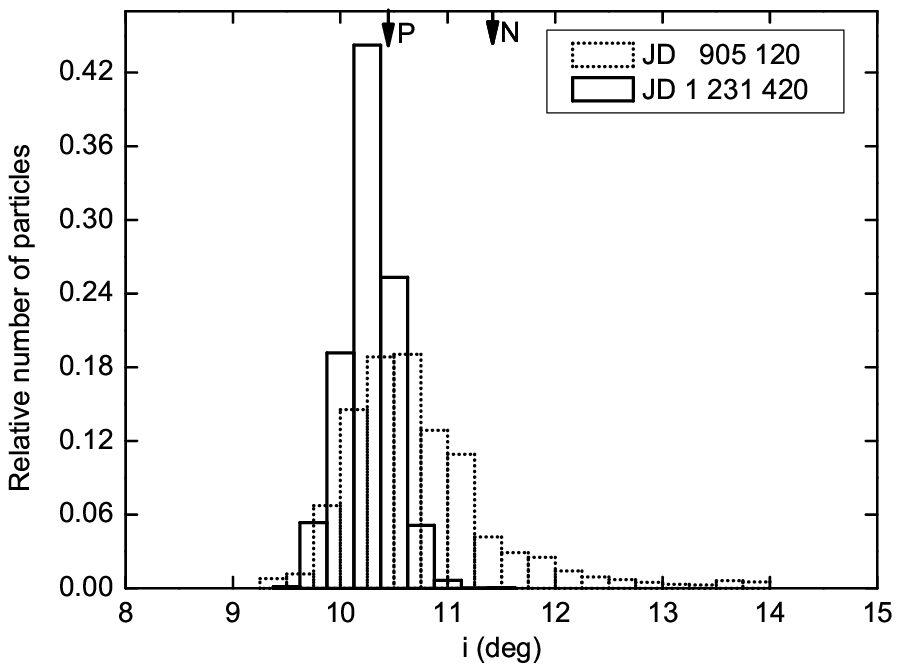}
    \hspace{0.3cm}
            \includegraphics[width=5.5cm,angle=0]{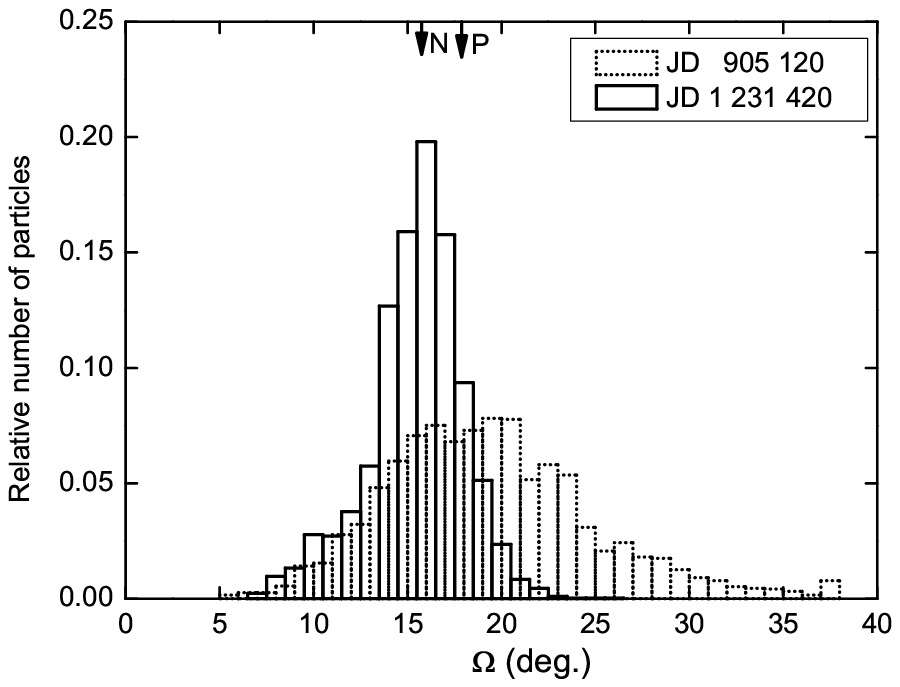}}
 \vspace{0.7cm}
\centerline{\includegraphics[width=4.5cm,angle=-90]{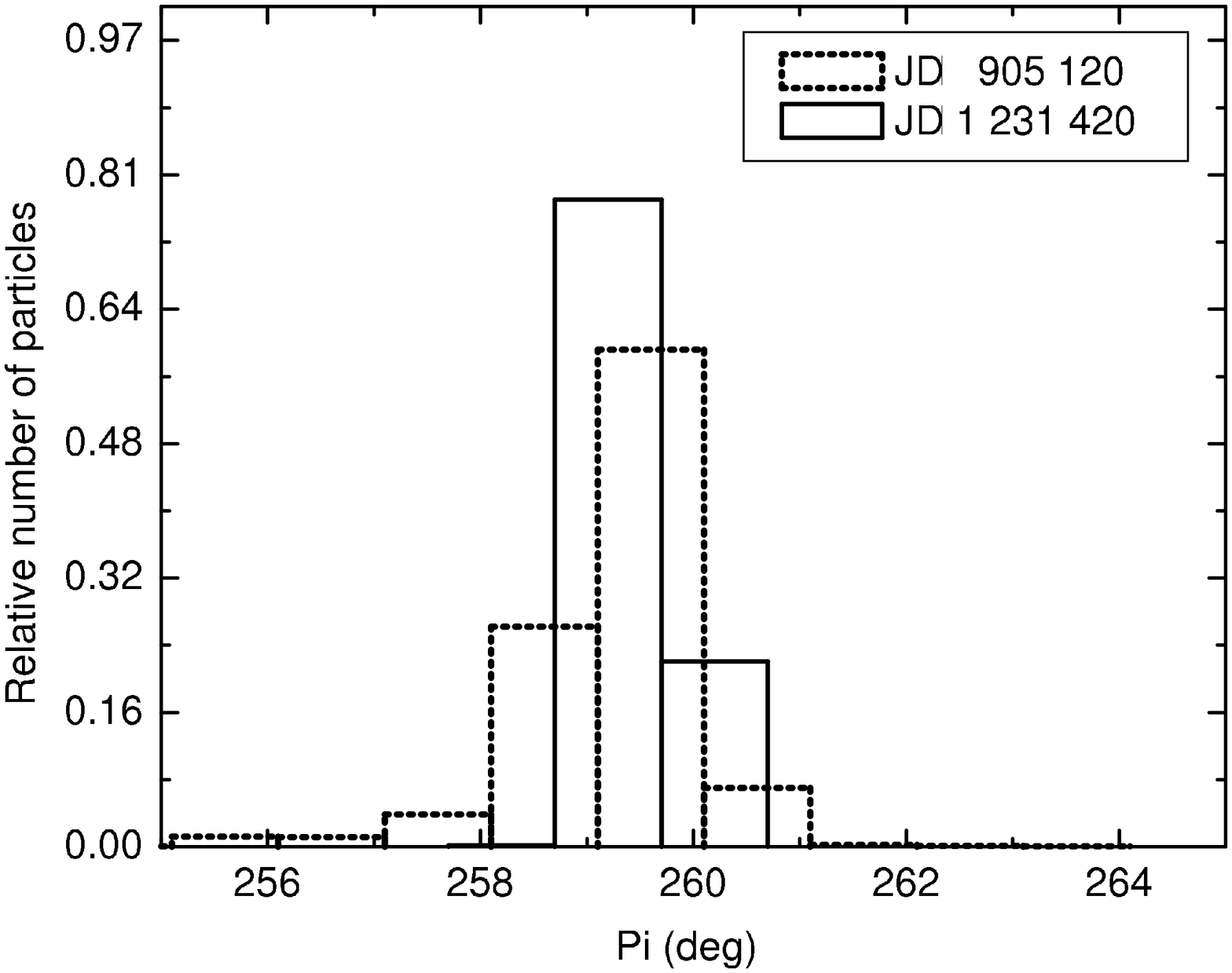}
    \hspace{0.3cm}
            \includegraphics[width=4.5cm,angle=-90]{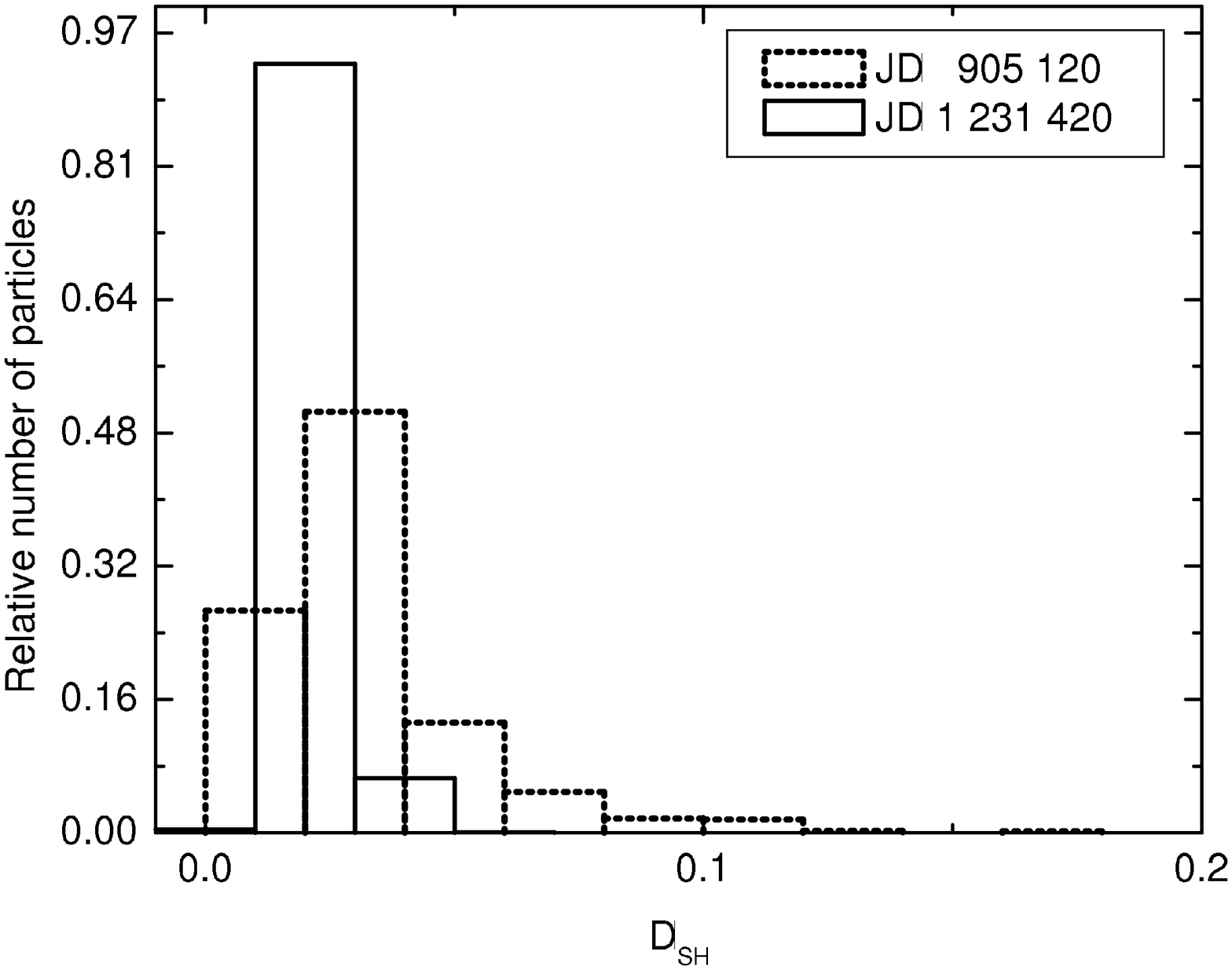}}
\caption{Histograms of orbital elements $a, e, i, \Omega, \pi$ and the D-criterion of
 modeled particles that left the parent body in the events $JD=905\,120$ and
 $JD=1\,231\,420$. $\downarrow$P and $\downarrow$N display the actual value of
 orbital element of P\v{r}\'{i}bram and Neuschwanstein.}
  \label{F4}
\end{figure*}

The total number of modeled particles does not provide sufficient
sample of meteors crossing the Earth's orbit during the activity
period. Therefore, we adopt the analogue of the target plane
defined as the plane going through the center of the Earth and
perpendicular to the velocity vector of the asteroid during its
close fly-by. The plain was placed to the position and orientation
at the moment of the P\v{r}\'{i}bram meteorite fall in 1959.
Target hits by released particles within one of their entire
revolutions around the Sun are displayed in Figure 5. Particles
released in the older event are spread to a larger area.

\begin{figure*}
\centerline{\includegraphics[width=6.5cm,angle=0]{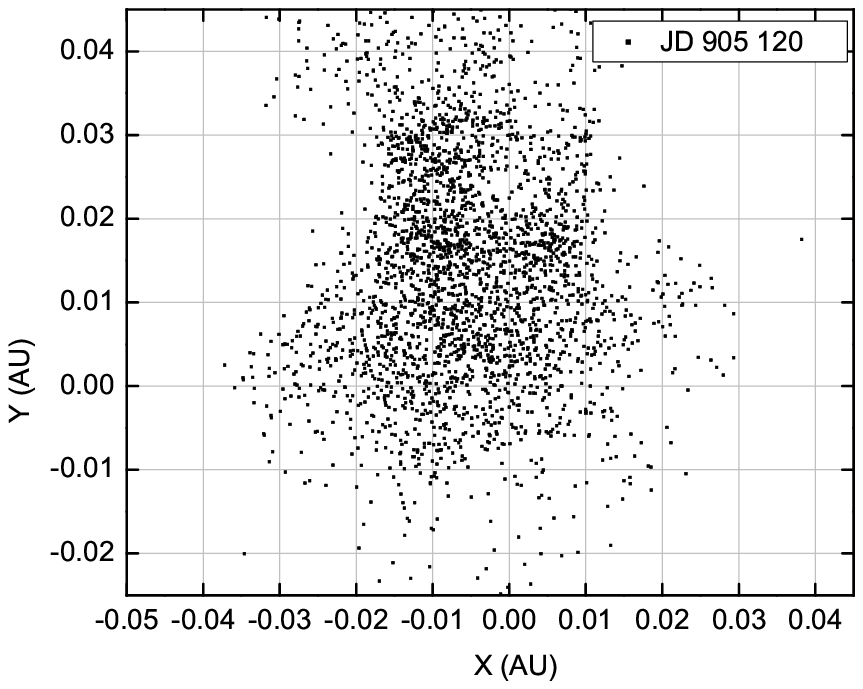}
            \hspace{0.3cm}
            \includegraphics[width=6.4cm,angle=0]{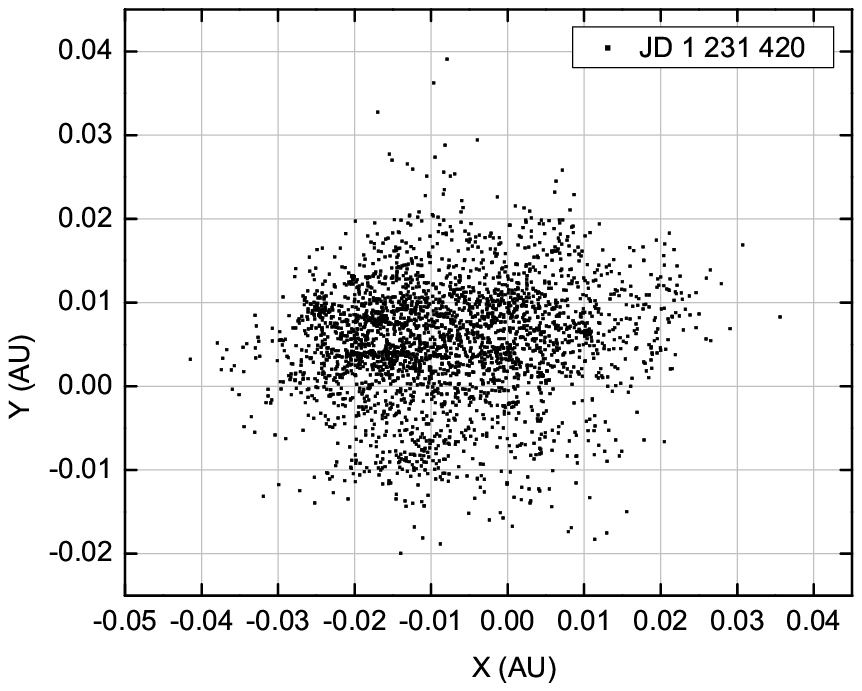}}
\caption{The cross-section  of the target plain at the Earth with hits by the model
 particles after they complete one revolution around the Sun.  Left - particles released
 in $JD=905\,120$, right - particles released in $JD=1\,231\,420$. The
 position (0,0) represents the Earth.}
   \label{F5}
\end{figure*}

From the geometry of the particle motion through the target plain
one can estimate that the Earth crosses the stream for approx. 8
days. Figure 6 shows the radiant positions of modeled particles
that hit the Earth in 2009. The radiation area of simulated
meteors is almost $5\degr \times 5\degr$ wide in RA and Dec. The
plot also displays the actual positions of P\v{r}\'{i}bram and
Neuschwanstein radiants. If the observed meteors from the
P\v{r}\'{i}bram stream come from the tidal disruption of the
parent asteroids in the event 4200 years ago, the real meteors
could have similar radiant distribution during the shower activity
as shown in Figure 6. The mean radiant has coordinates
$RA=192.8\degr\pm 1\degr$, $Dec=18.3\degr\pm 2\degr$ and the
ephemeris of the radians is then derived as:

\begin{equation}
\begin{array}{l}
\displaystyle RA=192.8\degr + 0.73~(L_{\odot}-17.79\degr)\\
\displaystyle DC=~18.3\degr + 0.51~(L_{\odot}-17.79\degr)
\end{array}
\label{eq:xdef}
\end{equation}

\noindent where 17.79\degr represents the Solar longitude of
P\v{r}\'{i}bram fall in 1959 (eq.\,2000.0).

\begin{figure}
\centerline{\includegraphics[width=6.2cm,angle=-90]{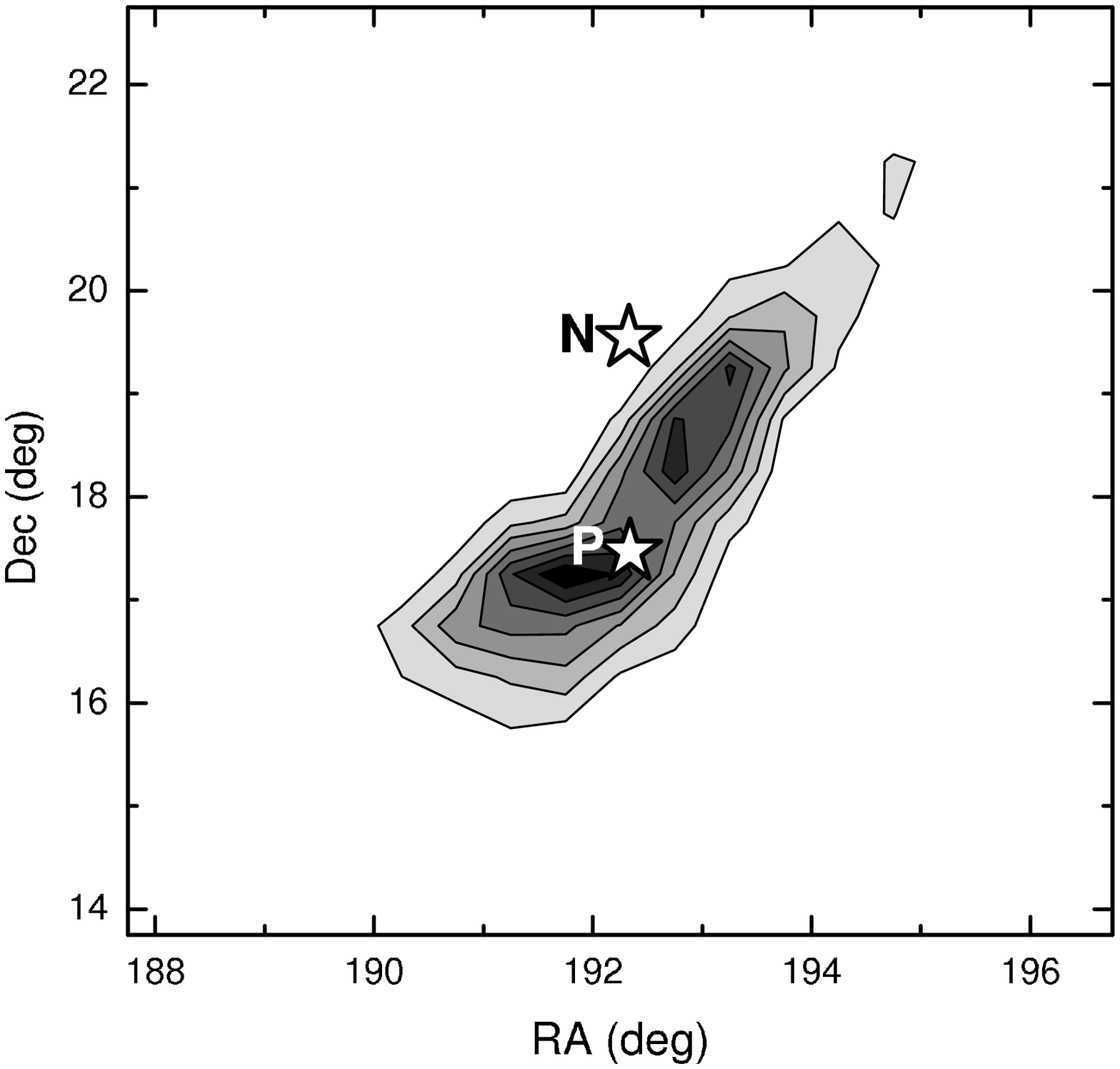}}
 \caption{Radiants of modeled meteors in 2009 (eq.\,2000.0). Contour plot
  shows a theoretical radiant probability density distribution if particles
  were released in the older event. Radiants of P\v{r}\'{i}bram and
  Neuschwanstein are marked.}
 \label{F6}
\end{figure}

To estimate the meteor activity of the prospective meteor stream
along the orbit of P\v{r}\'{i}bram, we need to know the size
distribution of released particles and their total number. As we
already mentioned above, the model asteroid has the same physical
properties as the near Earth asteroid Itokawa. Precise
measurements of its mass, density, topography, volume and porosity
are known from the in-situ exploration by Japanese probe Hayabusa
(\citealt{abe b}). The analysis of the detail surface images
provides data about the pebbles and boulders cumulative size
distribution (\citealt{sai}):

\begin{equation}\label{1}
N(>D)=BD^{-2.8}~,
\end{equation}

\noindent where the slope -2.8 was derived for the cumulative size
distribution for surface features with diameters in the range from
20\,cm to 20\,m. Supposing that the same size distribution is
valid for the wider interval of sizes between 0,01\,m and 30\,m,
the total mass of the asteroid is given as:

\begin{equation}\label{1}
M=- \int^{30}_{0.01} 2.8BD^{-3.8}m(D,\varrho)dD~,
\end{equation}

\noindent where $m(D,\varrho$) represents the mass of each pebble
with the diameter $D$ and the density $\varrho$.

Since the total mass and macroporosity (40\,\%) for Itokawa are
known (\citealt{abe b}), the density of the particles on the
Itokawa is approx. $\varrho=3,25\,g\,cm^{-3}$. Also the absorption
spectrum of Itokawa in the near-infrared channel shows features
similar to the LL5 or LL6 chondrite spectra (\citealt{abe a}),
with very similar densities $3.29 \pm 0.17\,g\,cm^{-3}$
(\citealt{wilk}). Generally, Itokawa-like body might be a
progenitor of P\v{r}\'{i}bram- and Neuschwanstein-like meteorites.

The constant $B$ is then simply derived from equation (3) and the
total number of all particles within the size range is calculated
according to equation (2). The main idea is that the total mass of
the asteroid will be disrupted to the boulders according to the
mentioned size distribution. That is why the parent body must be a
weakly bound rubble-pile asteroid. The equation (2) leads to about
4$\times10^6$ particles larger than $D\geq$ 0.6\,m or
3.8$\times10^{11}$ particles larger than $D\geq$ 1\,cm bound in
the entire volume of the asteroid.

If we spread the total number of particles with the same size
distribution along the orbit of the putative parent body, we can
estimate the activity of the shower, for instance for
one-centimeter particles and larger, as shown in Figure 7. The $y$
axis on the left gives the individual inflow of particles onto the
entire Earth in the one-day interval, the right $y$ axis gives the
biased activity for the one observing site (Central European
conditions, e.g. at the Astronomical and Geophysical Observatory
Modra with the all-sky video system, \citealt{toth}). The
observing site counts were biased due to the burning height of the
meteors over the surface (70\,km), the fraction of observed time
(night/day ratio in April, the month of P\v{r}\'{i}bram and
Neuschwanstein falls and proposed activity of the putative shower)
and the weather conditions on the observing site (long term
meteorological observations at AGO give 50\% chance for observing
at night).

\begin{figure}
\centerline{\includegraphics[width=8.0cm,angle=0]{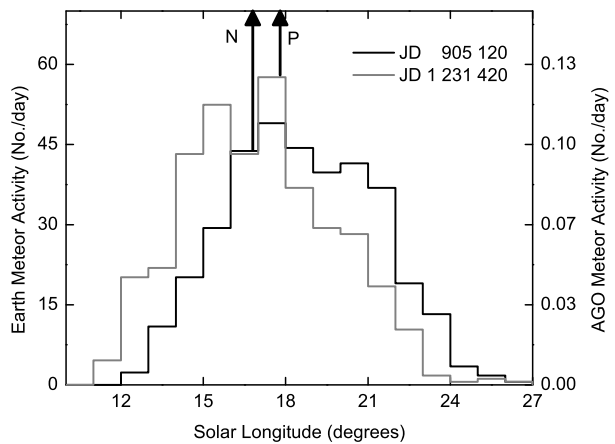}}
 \vspace{0.25cm} \caption{The activity of the potential meteor
  shower coming from the putative P\v{r}\'{i}bram meteorite
  progenitor. The left $y$ axis gives the number of particles that
  hit the Earth in the one-day interval, the right $y$ axis gives the
  number of meteors observed from the AGO Modra site including
  biased observation due to bad weather, visible meteoric area, night/day ratio.}
 \label{F7}
\end{figure}

\section{Frequency of the NEA break-up by Earth tides}

In this section we estimate the frequency of tidal disruption of
close approaching NEAs to the Earth. Considering a rubble-pile
structure of NEAs larger than $\sim$ 200 meters, the frequency of
such close approaches within the Roche limit of the Earth ($\sim 2
R_{\oplus}$ for a non rotating body) is one per 11\,000 years
based on the impact frequency of NEAs (\citealt{brown}).

\begin{figure}
 \centerline{\includegraphics[width=6.2cm,angle=-90]{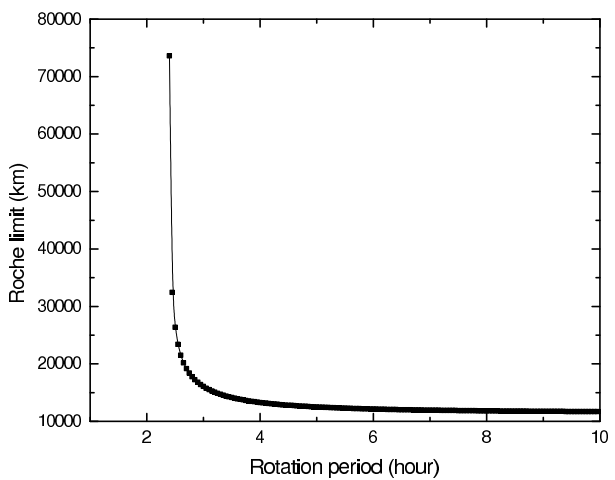}}
 \caption{The Roche limit of tidal disruption as a function of asteroid rotation period
  close to spin barrier.}
 \label{F8}
\end{figure}

\begin{figure}
\centerline{\includegraphics[width=6.2cm,angle=-90]{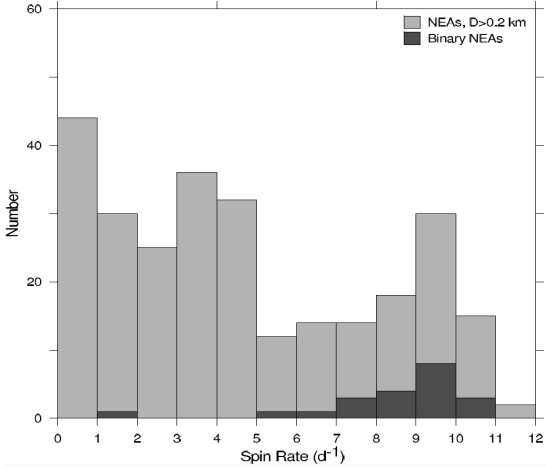}}
 \caption{Distribution of spin rate of NEAs larger than 200\,m (\citealt{pravec}). Spin
  barrier is around 11 periods per day.}
 \label{F9}
\end{figure}

Moreover, considering the rotation of the asteroid, the
centrifugal acceleration on the asteroid's equator shifts the
Roche limit of tidal break up farther from the Earth. The
centrifugal acceleration due to the motion of the asteroid around
the Earth (hyperbolic trajectory) was neglected. The function of
the Roche limit depending on the asteroid rotation period is
derived from the condition for the gravitational, centrifugal and
tidal acceleration affecting the particle on the surface of the
asteroid $a_g = a_{\omega}+a_{tid}$~:

\begin{equation}\label{1}
R_{Roche}=\sqrt[3]{\frac{GM_{\oplus}}{2\pi(\frac{G\rho}{3}-\frac{\pi}{P^{2}})}}~,
\end{equation}

\noindent where $R_{Roche}$ - the Roche limit depending on the
spin ratio, $G$ - gravity constant, $M_{\oplus}$ - mass of the
Earth, $\rho$ - bulk density of rubble pile NEA
(1900\,kg\,m$^{-3}$, same as Itokawa) and $P$ is an asteroid
principal axis rotation period. The function is depicted in Figure
8.

Therefore, the increase of the Roche limit due to asteroidal
rotation changes the frequency of tidal disruption of NEAs in a
close vicinity of the Earth. The frequency of Earth impacts
(\citealt{rich}) is given as

\begin{equation}\label{1}
F_{imp}= {P_{intr} N_{D} \pi R^{2}_{\oplus}~(1 +
\frac{v^{2}_{esc}(R_{\oplus})}{v^{2}_{\infty}})}~,
\end{equation}

\noindent where $P_{intr}$ is the intrinsic collisional
probability of Earth-crossing asteroids (ECAs) with the Earth,
$N_{D}$ is a cumulative number of ECAs larger than D (in our case
200\,m), $R_{\oplus}$ is the Earth radius  (\citealt{rich}). The
part enclosed in the parentheses expands the impact area due to
the Earth gravitational attraction.

The equation (5) is used for the estimation of the frequency of
the tidal breakups considering the rotation of the asteroid.
According to the spin rate distribution of NEAs larger than 200\,m
(\citealt{pravec}, Figure 9 in this paper), there are 28$\%$ of
bodies rotating in the range of spin rate 6 - 11 per day, near the
spin barrier $\sim$ 2.2 h. The Roche limit is substantially larger
within this spin rate interval. Assuming the impact frequency on
the Earth (\citealt{brown,wer}), we obtain a total frequency of
close approaches within the Roche limit as a function of asteroid
rotation. For each bin of the histogram in Figure 9, we can derive
the equation similar to (5) with the corresponding Roche limit
(4). Then the sum in the equation (6) means the total frequency of
the NEA occurrence within the extended impact target. It is
calculated from each interval $i$ of the spin rate distribution.

The ratio of tidal disruptions in close vicinity of the Earth
$F_{tid}$ to impact frequency of NEAs on the Earth $F_{imp}$ as
follows:

\begin{equation}\label{1}
\frac{F_{tid}}{F_{imp}}= \frac{\sum_{i} P_{intr} N_{D(i)} \pi R^{2}_{Roche(i)}~(1 +
\frac{v^{2}_{esc(i)}(R_{Roche(i)})}{v^{2}_{\infty}})}{P_{intr} N_{D} \pi
R^{2}_{\oplus}~(1 + \frac{v^{2}_{esc,\oplus}(R_{\oplus})}{v^{2}_{\infty}})}~,
\end{equation}

\noindent where $N_{D(i)}= N_{D}.w_{i}$ and $w_{i}$ is a fraction
of NEAs with given rotation period in i-th interval (Figure 9).

Using above calculation, the total frequency of tidal disruption
of NEAs is seven times higher compared to the impact frequency of
the entire NEA population larger than 200\,m. For the specific
impact frequency of 200\,m NEAs equal to $2.3\times10^{-5}$ per
year on the Earth (\citealt{ivan}), the tidal disruption frequency
would be $1.6\times10^{-4}$ per year. It means about once per
6\,200 years.

\section{Tidal disruption as the source of asteroidal meteoroids}

The size-frequency distribution of near-Earth population
originated by many complicated processes. It is generally accepted
that the population of NEA is not primordial and it has come from
other sources, mostly from the Main belt. The mean lifetime of
objects on NEA orbits is also relatively short when compared with
the lifetime of the Solar system (\citealt{bott2002}). Also the
size distribution of the current NEA population might not copy the
distribution of the Main belt progenitor population, mostly
because several effects could change specific parts of population
by a different rate (e.g. Yarkovsky effect, collisions between
asteroids, etc.). Currently, models of the NEA population
distributions are developed according to debiased observational
data from telescopic surveys, crater counts on the Moon, from
evolution models of the NEA population and impact frequency onto
the Earth derived from annual large bolide influx. The tidal
disruption of weak-bound asteroids flying-by planets inside the
Roche limit could, therefore, enhance the population of NEAs of
smaller diameters and deplete the population of larger bodies.

The mean lifetime of the asteroid in the NEA space depends on its
evolutional path, and if considering only sources of asteroidal
objects, the total mean lifetime is about $4\times10^{6}$ years
(\citealt{bott2002}). During that time, until the object is swept
away from the NEA space, certain fraction of objects of certain
diameters has a chance to fly by the planet within the Roche limit
and to be disrupted. Nevertheless, the disruption outcome depends
on several parameters: the encounter velocity, asteroid shape,
spin axis orientation and spin rate. We set the lower and upper
limits for the size of objects incoming inside the Roche limit.
The lower limit is given by the fact that objects 200\,m and
larger are not observed having the rotation faster than the spin
barrier, which suggests their rubble-pile structures
(\citealt{pravec00}). To set the upper limit of disrupted NEA in
the mentioned size range, we adopted the \citet{sb} cumulative
size distribution for the NEA population. The model was altered
according to the impact frequency that shows the size distribution
of objects below 2\,km has shallower slope (-1.7, \citealt{ivan}):

\begin{equation}
\begin{array}{l}
\displaystyle N(>D)=BD^{-1.7} ; D\subset\langle200\,m;2\,km) \\
\displaystyle N(>D)=BD^{-2.3} ; D\subset\langle2\,km;6.6\,km)
\end{array}
\label{eq:xdef}
\end{equation}

\noindent Considering the steady-state population of NEA, where
depleted objects are replenished from outer sources, we get the
upper limit for the asteroid size of 6.6 km. Such an object
encounters the Earth below the Roche limit at most once during its
lifetime in the NEA space (4$\times10^6$).

\begin {figure}
\centerline{\includegraphics[width=8.0cm,angle=0]{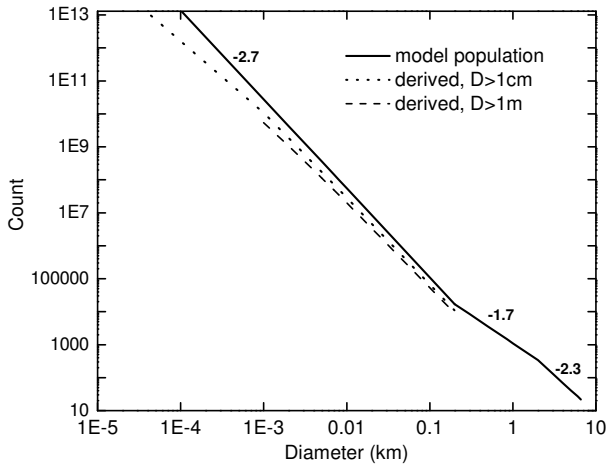}}
  \vspace{0.25cm} \caption{Cumulative size distribution of the NEA
   population according to \citet{wer} with the slopes -2.3 and
   -1.7 and by \citet{brown} with the slope -2.7. The dotted curve represents the
   population without the contribution of the tidal disrupted matter if
   resulting particles of disruptions are larger than 1\,m (slope -2.45),
   the dashed curve for particles larger than 1\,cm (slope of the population -2.55).}
 \label{F10}
\end{figure}

We have found that at most $4.2\pm0.5\%$ of the NEA population
undergoes tidal disruption due to Earth encounter within its
lifetime assuming that each body that flies by the Earth within
the Roche limit will be disrupted since it is possible that some
fraction of the disrupted matter could be reaccumulated
(\citealt{rich}). The total mass loss in the size range from
200\,m to 6.6\,km per $4\times10^{6}$ years is approx.
$10^{15}\,kg$. If we assume that all this mass is redistributed in
a new cumulative size distribution with the slope equal to -2.8
(the slope of boulders and pebbles on Itokawa) within the size
range from 1\,cm to 200\,m then the primordial slope of the
population of NEA smaller than 200\,m will be enhanced with the
steeper population coming from the tidal disruption. There are
several estimates of NEA population size distribution within the
meteoroid sizes (1\,cm-200\,m), e.g. \citet{ivan} and
\citet{brown}. If we adopt the model derived from the bolide
counts with the slope equal to -2.7 (\citealt{brown}) and subtract
the derived population from the tidal disruption, we reveal the
size distribution of the NEA population as it would look like
without the contribution from the tidal disintegration from the
Earth. With respect to how finely the matter will be disrupted
(down to 1\,cm or 1\,m), the resulting slope of the primordial
population created by all effects, but not by tidal disruption,
would be -2.55 or -2.45. It seems that the tidal disruption of the
larger rubble-pile asteroids may change the size distribution
slope of the population coming to the NEA region and makes it
steeper.

\section{Conclusion}

This work concerns about the dynamical evolution and the activity
of the theoretical stream of meteoroids that originated by the
tidal disruption of the rubble-pile-like parent asteroid that we
assumed moved in the orbit of the P\v{r}\'{i}bram meteorite. The
disruption of the asteroid and creation of the stream emerged when
the node of the parent body got close to or crossed the Earth
orbit. In the last 5000 years, the orbit of P\v{r}\'{i}bram has
fulfilled this condition twice: 4200 and 3300 years ago. In both
events, the release of particles from the parent surface was
modeled, and their following orbital evolution until the current
date was traced. It takes several hundred years to spread
particles around the entire orbit of the parent asteroid.

We assumed that the parent asteroid was similar to the Itokawa the
complete disintegration of which would deliberate
3.8$\times10^{11}$ meteoroid particles with diameters larger than
$D\geq$ 1\,cm. The activity of the proposed meteor shower
observable in April is low according to Figure 7. Annual
observation campaigns are needed in the first two weeks of April
to detect any meteors coming from the proposed source.

The reverse approach to the problem might reveal what size and
what amount of the matter coming from the tidal disruption of the
asteroid were needed to create events like the P\v{r}\'{i}bram and
Neuschwanstein meteorite falls. If the average period between two
falls of meteorites is 43 years (as in the case of the meteorites
mentioned above), according to equation (2), such a stream of
meteoroids must contain 2.2$\times10^7$ - 1.0$\times10^{10}$
boulders of 0.6\,m diameter, which was expected pre-atmospheric
diameters of the meteorites (\citealt{revelle}). This number is in
a good agreement with the previous work (\citealt{spur}). The
parent body diameter can be estimated to be 0.6 - 5\,km according
to equation (3).

We have also estimated that the tidal disruption of rubble-pile
asteroids by the Earth tides might populate the NEA space with
smaller and compact meteoroids and, therefore, change the size
distribution with lifting its slope.

\section*{Acknowledgements} This work was supported by Slovak grant
agency VEGA, No. 1/0636/09 and grant of Comenius University No.
UK/245/2010.

\end{document}